\documentclass{PoS}

\usepackage{graphicx}
\usepackage{wrapfig}
\usepackage{url}
\def\aj{AJ}%
\def\apj{ApJ}%
\def\apjs{ApJS}%
\def\aap{A\&A}%
\def\pasa{PASA}%
\title{RadioAstron Early Science Program Space-VLBI AGN survey: 
strategy and first results}

\ShortTitle{RadioAstron Space-VLBI AGN survey}

\author{\speaker{Kirill V. Sokolovsky}\\
        Astro Space Center, Lebedev Physical Inst. RAS, Profsoyuznaya 84/32, 117997 Moscow, Russia\\
        Sternberg Astronomical Institute, Moscow University, Universitetsky 13, 119991 Moscow, Russia\\
        E-mail: \email{kirx@scan.sai.msu.ru}}
\author{for the RadioAstron AGN Early Science Working Group}

\abstract{
RadioAstron is a project to use the 10\,m antenna on board the dedicated
SPEKTR-R spacecraft, launched on 2011 July 18, to perform 
Very Long Baseline Interferometry from space -- Space-VLBI. 
We describe the strategy and highlight the first results of a
92/18/6/1.35\,cm fringe survey of some 
of the brighter radio-loud Active Galactic Nuclei (AGN) 
at baselines up to 25~Earth diameters ($D_{\oplus}$).
The survey goals include a search for extreme brightness
temperatures 
to resolve the Doppler factor crisis and to constrain possible 
mechanisms of AGN radio emission, studying the
observed size distribution of the most compact features in AGN radio
jets (with implications for their intrinsic structure and the
properties of the scattering interstellar medium in our Galaxy) and
selecting promising objects for detailed follow-up observations,
including Space-VLBI imaging.  Our survey target selection is based on 
the results of correlated visibility measurements at the longest ground-ground baselines 
from previous VLBI surveys. The current long-baseline fringe
detections with RadioAstron include OJ~287 at 10~$D_{\oplus}$ (18\,cm), BL~Lac at
10~$D_{\oplus}$ (6\,cm) and B0748$+$126 at 4.3~$D_{\oplus}$ (1.3\,cm).
The 18 and 6\,cm-band fringe detections at 10~$D_{\oplus}$ imply brightness
temperatures of $T_b \sim 10^{13}$\,K, about two orders of magnitude above the
equipartition inverse Compton limit.  These high values of $T_b$ 
might indicate that the jet flow speed is often higher than the jet
pattern speed.}

\FullConference{11th European VLBI Network Symposium \& Users Meeting,\\
		October 9-12, 2012\\
		Bordeaux, France}

\begin{document}

\section{The space radio telescope}

\begin{figure}[htb]
\centering
\includegraphics[height=6.8cm]{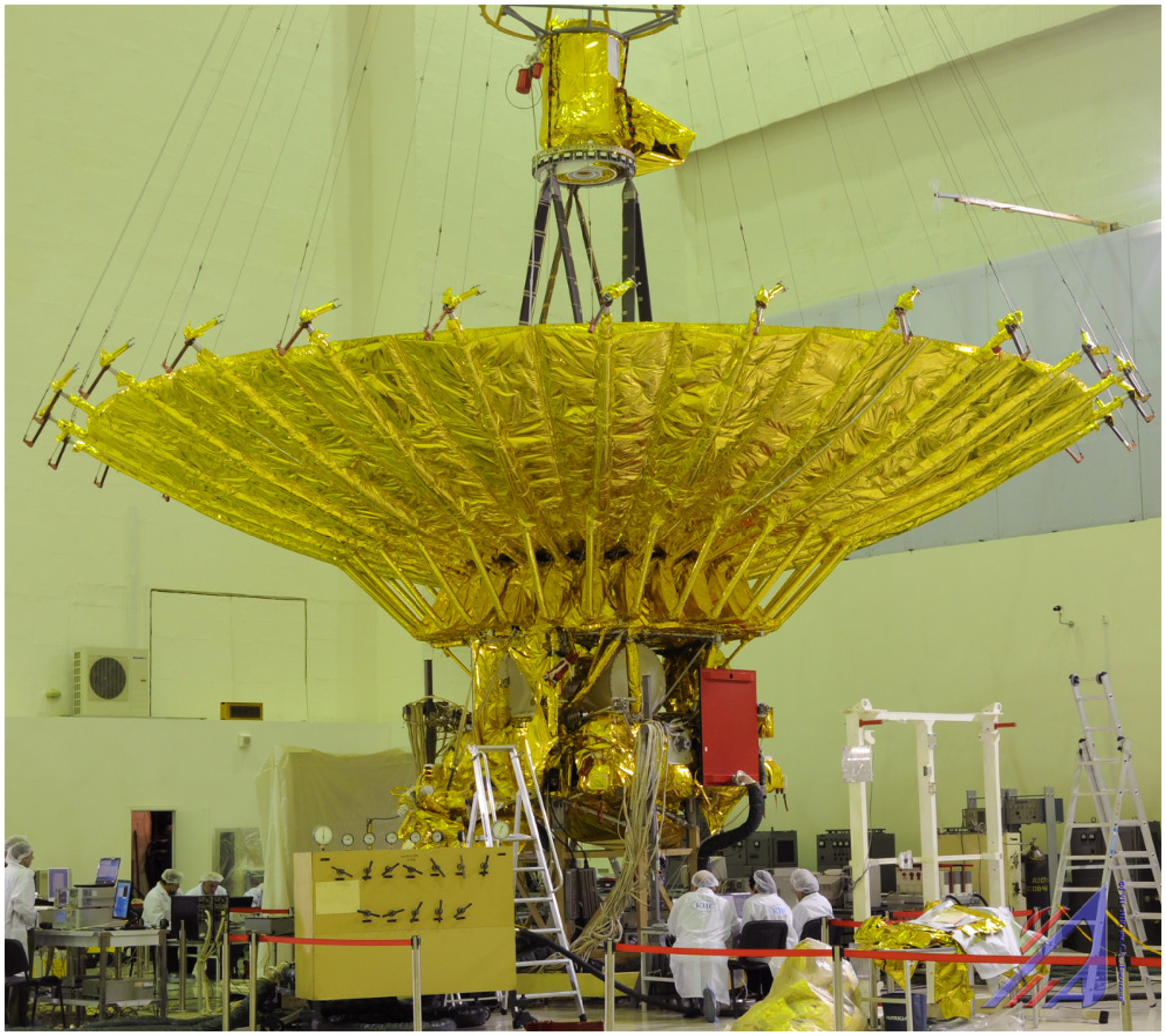}~~
\includegraphics[height=6.8cm]{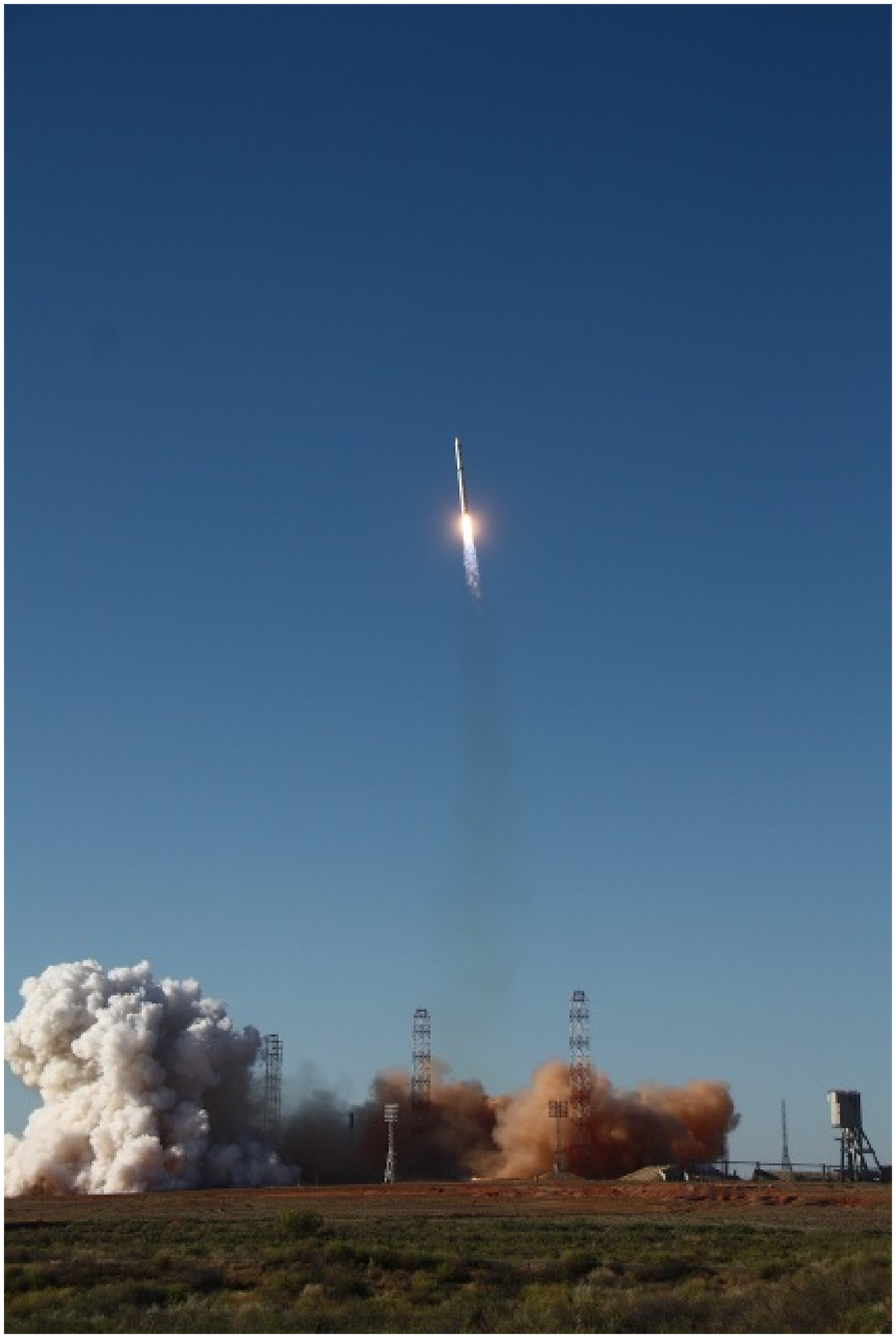}
\caption{SPEKTR-R assembled at Lavochkin Association (left) and its launch
from Baikonur (right).}
\label{fig:SRTLavochkin}
\end{figure}

The 10\,m space radio telescope of the RadioAstron project is installed on
board of the dedicated SPEKTR-R spacecraft. The spacecraft was launched into a
highly elliptical orbit on 2011 July~18 from the Baikonur Cosmodrome by a
Zenit-3F rocket (Figure~\ref{fig:SRTLavochkin}).
The orbit was selected so that its parameters evolve under the gravitational pull
of the Moon, to provide a wide range of projected baselines for VLBI
observations of various sky regions during the mission lifespan.
As of 2012 October~2, the SPEKTR-R orbital parameters were:
206-hour period, 73\,000\,km perigee, 281\,000\,km apogee, and 79$^\circ$ inclination.
Regular measurements of the spacecraft's distance and velocity using
standard radiometric techniques supported by laser ranging, direct optical 
imaging and VLBI state vector measurements \cite{2012cosp...39..696G} 
allow one to reconstruct its position and velocity with accuracies 
typically better than $\pm500$\,m and $\pm2$\,cm~s$^{-1}$, respectively.

The space telescope is equipped with 92\,cm (324\,MHz, P-band), 18\,cm
(1.7,GHz, L-band), 6\,cm (4.8\,GHz, C-band) and
1.3\,cm (22.2\,GHz, K-band) receivers, an on-board hydrogen maser, and a high-gain antenna system
to downlink VLBI data to a ground station in real time. Currently, the 22\,m
antenna of the Pushchino radio astronomy observatory near Moscow, Russia serves as
the ground data acquisition station. The second RadioAstron data acquisition station is
under construction on the basis of the National Radio Astronomy
Observatory's 43\,m telescope at Green Bank, West Virginia, USA.

\section{AGN survey strategy}

A fringe detection survey of radio-bright AGN is
being conducted as part of the RadioAstron Early Science Program. The survey
goals include a search for extreme brightness temperatures, $T_b$, 
to resolve the Doppler crisis \cite{2006ApJ...640..185H} and to constrain
possible mechanisms of AGN radio emission \cite{2002PASA...19...77K},
to study the size distribution 
of the most compact features in AGN
radio jets (with implications for their intrinsic structure and the properties of
the scattering in the interstellar medium in our Galaxy) and select promising targets 
for detailed follow-up observations, including space-VLBI imaging.

The survey target selection is based on the results of the correlated visibility 
measurements on the longest ground-ground baselines from existing VLBI surveys 
including the 13/3.6\,cm (S/X-band) VLBA Calibrator Surveys (VCS) 1 to 6
\cite{2002ApJS..141...13B,2003AJ....126.2562F,2005AJ....129.1163P,2006AJ....131.1872P,2007AJ....133.1236K,2008AJ....136..580P} 
and the Research and Development VLBA program (RDV)
\cite{1996ApJS..105..299F,1997ApJS..111...95F,2009JGeod..83..859P,2012A&A...544A..34P}, 
2\,cm (K$_{\rm u}$-band) observations of the MOJAVE
program\footnote{\url{http://www.physics.purdue.edu/astro/MOJAVE/}},
7\,mm (Q-) and 3\,mm (W-band) results of the Boston University
group\footnote{\url{http://www.bu.edu/blazars/}} and
of a Global 86\,GHz VLBI Survey of Compact Radio Sources~\cite{2008AJ....136..159L}, respectively.
We also consulted the list of high--$T_b$ sources observed
at 6\,cm by the VLBI Space Observatory Programme (VSOP)
\cite{2004ApJS..155...33S,2004ApJ...616..110H,2008ApJS..175..314D}.

The results of many ground-based VLBI surveys are summarized in the Radio Fundamental
Catalog\footnote{\url{http://astrogeo.org/rfc}}. We used the ``unresolved X-band flux
density'' 
parameter listed in this catalog to set scheduling priorities for the sources.
Among sources with comparable unresolved flux densities that satisfy 
RadioAstron visibility constraints, preference is given to sources for
which {\it (i)} the correlated visibility at 13, 3.6, 2\,cm and 7\,mm bands measured from
the ground is not decreasing rapidly with increasing baseline length,
{\it (ii)} we can obtain both short ($<5 D_{\oplus}$) and long projected space--ground
baselines within one or a few SPEKTR-R orbital revolutions,
{\it (iii)} VSOP measured $T_b>10^{12}$\,K, and
{\it (iv)} there is a 3\,mm detection. We try to schedule the preferred
sources first to maximize the detection rate in the early stages
of the survey.

\begin{wrapfigure}{r}{0.5\textwidth}
\centering
\includegraphics[height=0.48\textwidth,angle=270,clip,trim=0.1cm 0.1cm 0.05cm 0.05cm]{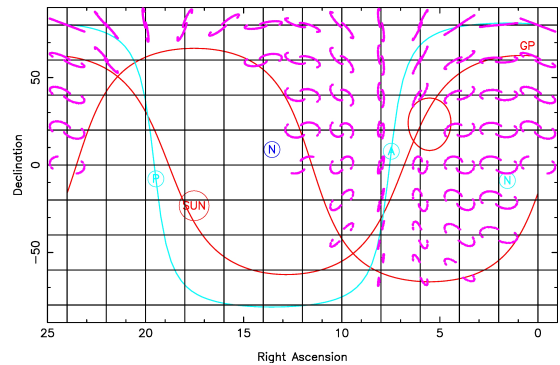}
\caption{Possible space-ground $(u, v)$-coverage for various sky positions
computed for 206~hours, one orbital revolution, starting on 2012 December~15 00:00~UT
with the telescopes: RadioAstron, EVN, LBA, Arecibo, GBT, Usuda.
Also marked on the plot are: the Galactic Plane (GP), Sun avoidance regions,
the satellite's
orbital plane, the perigee (P) apogee (A), and the direction perpendicular to the
orbital plane.}
\label{fig:allskyuv}
\end{wrapfigure}

All four RadioAstron bands (92, 18, 6, and 1.3\,cm) are employed in the survey, 
with the main focus on 18, 6, and 1.3\,cm bands. Most observations are done in 
a dual-band mode: 6$+$18\,cm or 6$+$1.3\,cm. The space radio telescope observes
simultaneously at two bands while the ground telescopes are divided in two sub-arrays
or switch between the two bands during the experiment. A typical AGN survey
observation consists of four scans, each 10-minute long, on a target source,
separated from other series of scans by the 40--60~minutes necessary to satisfy
the spacecraft's thermal constraints and slew to the next target.
The main factors affecting the scheduling include: a
Sun avoidance angle of 90$^\circ$ plus a small solid angle centered on anti-solar direction, 
satellite visibility to the tracking station (TS), TS visibility to the
satellite's high-gain antenna, target source visibility for ground
telescopes, and availability of ground telescopes during the time period when
all the other constraints are met. An example of the $(u, v)$-coverage computed with the
above constraints, except the last one, over the whole sky is presented on
Figure~\ref{fig:allskyuv}, although it should be noted that the plot
represents a period of rather favorable observing conditions.

Ground telescopes participating in the AGN Early Science Program
observations include the
Arecibo 300\,m and NRAO GBT 100\,m  (USA), 
ATCA tied-array of 5x22\,m, Parkes 64\,m, Mopra 22\,m, Hobart 26\,m, Tidbinbilla 70\,m (Australia),
Effelsberg 100\,m (Germany), 
Evpatoria 70\,m (Ukraine), 
Hartebeesthoek 26\,m (South Africa),
Jodrell Bank 70\,m (UK),
Medicina 32\,m and Noto 32\,m (Italy),
Shanghai 25\,m and Urumqi 25\,m (China),
Svetloe 32\,m, Zelenchukskaya 32\,m, Badary 32\,m (Russia),
Torun 32\,m (Poland),
Usuda 64\,m (Japan),
WSRT 14x25\,m (Netherlands),
Yebes 40\,m and Robledo 70\,m (Spain), as well as the EVN, Kvazar-KVO, and LBA arrays.

\section{First results}

\begin{wrapfigure}{r}{0.5\textwidth}
\centering
\includegraphics[width=0.48\textwidth,angle=0,clip,trim=4cm 1cm 2cm 4cm]{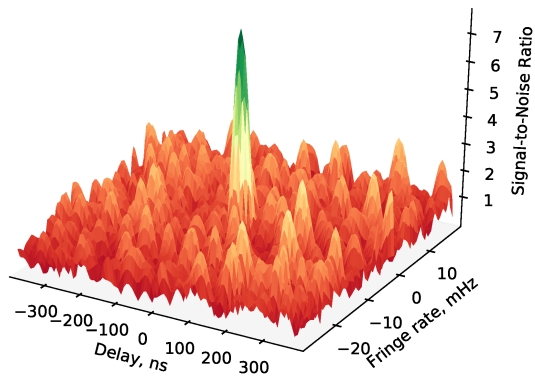}
\caption{Interferometric signal from the quasar B0748$+$126 detected between RadioAstron and
GBT (1.3\,cm) at the projected baseline of $4.3 \, D_{\oplus}$. The plot shows
the signal to noise ratio as a function of residual delay and rate after
delay-model subtraction and fringe fitting.}
\label{fig:fring}
\end{wrapfigure}

While the first months of the survey were marked by continuing development
of correlation techniques, choosing optimal space telescope observing modes,
and debugging the satellite VLBI data downlink system, the observations
during this period nonetheless
provided some record-breaking results. Fringes to the space telescope 
at 18 and 6\,cm bands were detected on projected baselines of about 10~$D_{\oplus}$ 
for blazars B0748$+$126, OJ~287, and BL~Lacertae. 
The ground array consisted of Arecibo, GBT, and Effelsberg telescopes.
At 1.3\,cm, fringes between the space telescope and the GBT were detected for
B0748$+$126 at projected baselines up to 4.3~$D_{\oplus}$ (Fig.~\ref{fig:fring}). 
Some of the current long-baseline fringe 
detections are presented in Table~\ref{tab:results}.
Both B0748$+$126 and OJ~287 exhibit interstellar
scintillation as determined in the MASIV Survey \cite{2008ApJ...689..108L};
BL~Lac was not included in MASIV.

\begin{table}
\begin{center}
\caption{Some of RadioAstron long-baseline detections as of January 2013}
\label{tab:results}
\begin{tabular}{ c c | c r r c l }
\hline
\hline
      &        &  Ground        & $B_{\rm max}$       & $B_{\rm max}/\lambda$ & $\lambda/B_{\rm max}$ & ~~~$z$ \\
Name  & Alias  &  telescope     &  ($D_{\oplus}$) &  (M$\lambda$)~~   & (mas)             &        \\
\hline
 &           & \multicolumn{5}{c}{L-band ($\lambda = 18$\,cm)}  \\
J0854$+$2006 & OJ~287     & Arecibo 300\,m   & 10   & 708  &  0.29 & $0.306$ \\
 &          & \multicolumn{5}{c}{C-band ($\lambda = 6$\,cm)}  \\
J2202$+$4216 & BL~Lac     & Effelsberg 100\,m & 10 & 2124 & 0.10 & $0.0686$ \\
 &          & \multicolumn{5} {c}{K-band ($\lambda = 1.3$\,cm)} \\
J0750$+$1231 & B0748$+$126  & GBT 100\,m      &  4.3 & 4215 & 0.05 & $0.889$ \\
\hline
\end{tabular}
\end{center}

{\bf Column designation:} (1)~source name and its alias~(2), (3)~ground telescope, 
(4)~and~(5) -- maximum baseline at which fringes were found, (6)~angular scale, (6)~redshift
(MOJAVE database).

\end{table}

While the 92\,cm band is actively used for RadioAstron observations of pulsars, 
no 92\,cm space--ground fringes on AGN have been detected (few attempts so far).
This band was given a low priority because
interstellar scattering is most prominent at longer wavelengths and the 
angular resolution is lower than that of other RadioAstron bands. However, 
detection of 18\,cm fringes at 10~$D_{\oplus}$ suggests that
interstellar scattering might not always prevent low-frequency fringe detections
at long baselines and more 92\,cm AGN observations should be attempted.

\section{Summary}

The RadioAstron space interferometer is exploring angular scales that were
never before accessible at centimeter wavelengths.
The 18 and 6\,cm fringe detections at 10~$D_{\oplus}$ imply brightness temperatures 
$T_b \sim 10^{13}$\,K, about two orders of magnitude above the equipartition inverse Compton limit
of a~few~$\times 10^{11}$\,K \cite{1994ApJ...426...51R}. 
These brightness temperature values
may be reconciled with the standard $e^-$/$e^+$ incoherent synchrotron
radiation model if the emission is Doppler--boosted by a factor of
$\delta \equiv [\Gamma (1-\beta \cos \theta)]^{-1} \sim 100$, where
$\Gamma$ is the bulk Lorentz factor, $\beta$ the velocity in units of $c$
of the emitting plasma, and $\theta$ is the angle between the plasma flow
direction and the line of sight. The large values of $\delta$ derived from
the RadioAstron $T_b$ measurements combined with the equipartition inverse Compton limit 
argument are inconsistent with the typical values of $\delta$ derived from
ground-based VLBI kinematic data \cite{2007ApJ...658..232C,2009AJ....138.1874L}. 
This inconsistency might indicate that the jet flow speed is often higher than the jet pattern speed.
More observations at long baselines are planned to probe the range of
$T_b \sim 10^{14\mathrm{-}15}$\,K.
Surprisingly, interstellar scattering is not preventing fringe detection at
long baselines even at 18\,cm.

\section*{Acknowledgements}

The RadioAstron Space Radio Telescope was build and is operated by Lavochkin
Association and Astro Space Center in collaboration with Russian and
international institutions. 
The RadioAstron AGN Early Science Working Group is deeply grateful to the observers
and technicians at Arecibo, ATCA, Effelsberg, Evpatoria, GBT,
Hartebeesthoek, Hobart, Jodrell Bank, Medicina, Noto, Mopra, Parkes,
Shanghai, Urumqi, Svetloe, Zelenchukskaya, Badary, Torun, Tidbinbilla, Usuda, WSRT,
Yebes, and Robledo observatories for making this project possible.

This research was supported by the Russian Foundation for Basic Research 
(projects 11-02-00368 and 12-02-33101), the basic research program
``Active processes in galactic and extragalactic objects'' of the Physical 
Sciences Division of the Russian Academy of Sciences, and the Ministry of 
Education and Science of the Russian Federation (agreement No.~8405).

\end{document}